\documentclass[lettersize,journal]{IEEEtran}
\usepackage{algorithm,amsbsy,amsmath,amssymb,epsfig,bbm,mathrsfs,multirow,amsthm}
\usepackage{color}
\usepackage{algpseudocode,tabularx}
\usepackage{graphicx,caption,subcaption,array,multirow,bm}
\usepackage{makecell}
\usepackage{cite}
\usepackage{threeparttable}

\newcolumntype{M}[1]{>{\centering\arraybackslash}m{#1}}
\newcolumntype{P}[1]{>{\hspace{0pt}}p{#1}}

\DeclareMathAlphabet{\mathpzc}{OT1}{pzc}{m}{it}

\makeatletter
\def\endthebibliography{%
\def\@noitemerr{\@latex@warning{Empty `thebibliography' environment}}%
\endlist
}
\makeatother
\makeatletter
\newcommand{\multiline}[1]{%
\begin{tabularx}{\dimexpr\linewidth-\ALG@thistlm}[t]{@{}X@{}}
#1
\end{tabularx}
}
\makeatother

\begin{document}

\title{Dynamic Spectrum Access for Ambient Backscatter Communication-assisted D2D Systems with Quantum Reinforcement Learning}

\author{Nguyen Van Huynh,~\IEEEmembership{Member,~IEEE}, Bolun Zhang, Dinh-Hieu Tran,~\IEEEmembership{Member,~IEEE}, Dinh Thai Hoang,~\IEEEmembership{Senior Member,~IEEE}, Diep N. Nguyen,~\IEEEmembership{Senior Member,~IEEE}, Gan Zheng,~\IEEEmembership{Fellow,~IEEE}, Dusit Niyato,~\IEEEmembership{Fellow,~IEEE}, and Quoc-Viet Pham,~\IEEEmembership{Senior Member,~IEEE}
\thanks{Nguyen Van Huynh is with the Department of Electrical Engineering and Electronics, University of Liverpool, Liverpool, L69 3GJ, United Kingdom (e-mail: huynh.nguyen@liverpool.ac.uk).}
\thanks{Bolun Zhang is with the School of Informatics, University of Edinburgh, Edinburgh EH8 9AB, United Kingdom (e-mail: s2145535@ed.ac.uk).}
\thanks{Gan Zheng is with the School of Engineering, University of Warwick, Coventry, CV4 7AL, UK (e-mail: gan.zheng@warwick.ac.uk).}
\thanks{Dinh-Hieu Tran is with the Interdisciplinary Centre for Security, Reliability and Trust (SnT), the University of Luxembourg, 1855 Luxembourg City, Luxembourg (e-mail: hieu.tran-dinh@uni.lu).}
\thanks{Dinh Thai Hoang and Diep N. Nguyen are with the School of Electrical and Data Engineering, University of Technology Sydney, Sydney NSW 2007, Australia (e-mail: hoang.dinh@uts.edu.au; diep.nguyen@uts.edu.au).}
\thanks{Dusit Niyato is with the College of	Computing and Data Science, Nanyang Technological University, Singapore	639798 (e-mail: dniyato@ntu.edu.sg).}
\thanks{Quoc-Viet Pham is with the School of Computer Science and Statistics, Trinity College Dublin, The University of Dublin, Dublin 2, D02 PN40, Ireland (e-mail: viet.pham@tcd.ie).}
}

\maketitle

\begin{abstract}
Spectrum access is an essential problem in device-to-device (D2D) communications. However, with the recent growth in the number of mobile devices, the wireless spectrum is becoming scarce, resulting in low spectral efficiency for D2D communications. To address this problem, this paper aims to integrate the ambient backscatter communication technology into D2D devices to allow them to backscatter ambient RF signals to transmit their data when the shared spectrum is occupied by mobile users. To obtain the optimal spectrum access policy, i.e., stay idle or access the shared spectrum and perform active transmissions or backscattering ambient RF signals for transmissions, to maximize the average throughput for D2D users, deep reinforcement learning (DRL) can be adopted. However, DRL-based solutions may require long training time due to the curse of dimensionality issue as well as complex deep neural network architectures. For that, we develop a novel quantum reinforcement learning (RL) algorithm that can achieve a faster convergence rate with fewer training parameters compared to DRL thanks to the quantum superposition and quantum entanglement principles. Specifically, instead of using conventional deep neural networks, the proposed quantum RL algorithm uses a parametrized quantum circuit to approximate an optimal policy. Extensive simulations then demonstrate that the proposed solution not only can significantly improve the average throughput of D2D devices when the shared spectrum is busy but also can achieve much better performance in terms of convergence rate and learning complexity compared to existing DRL-based methods.
\end{abstract}


\section{Introduction}
With the rapid development of wireless technologies, it is expected that future communication systems, e.g., 5G Advanced/6G, will need to support an enormous number of heterogeneous wireless devices. Many of these wireless devices require short-range, high-rate, and low-latency communications. To accommodate these requirements and increase spectrum efficiency, device-to-device (D2D) communication has been proposed. In particular, D2D communication enables communication between adjacent wireless devices without the use of network infrastructures like base stations (BSs)~\cite{Asadi2014Survey}. As such, D2D communication can enable direct communications between devices and reduce end-to-end latency. A key advantage of the D2D communication technology is that it can reuse the spectrum of cellular systems for direct communications between D2D devices, resulting in better spectrum and energy efficiency~\cite{Jiang2016Energy}. With these features, D2D communication is expected to be an essential part of 5G Advanced and 6G, especially in IoT and vehicular communications~\cite{Ansari20185G}. However, D2D communication may introduce interference to cellular users (CUs) if it reuses the same spectrum which the CUs are currently occupying. This is particularly challenging in future wireless networks where a large number of wireless devices operate in a dense area. For that, there is a demand for effective spectrum access strategies for D2D devices when opportunistically reusing the cellular spectrum.

\subsection{Related Work}
Numerous methods have been proposed in the literature to enable dynamic/opportunistic spectrum access to D2D communications. For example, the authors in~\cite{Lin2014Spectrum} considered the spectrum sharing problem for D2D communications in cellular networks. By considering both the overlay and underlay modes, the authors aimed to obtain an optimal spectrum sharing strategy that enables D2D devices to orthogonally share the spectrum with CUs or opportunistically use the frequency/time resources occupied by CUs. Then, analytical rate expressions were derived to optimize the two spectrum sharing modes based on a weighted proportional fair utility function. Similarly, the authors in~\cite{Chu2019Opportunistic} optimized the spectrum sharing for D2D-based ultra-reliable low latency communications. A rate optimization problem for the considered D2D network was first formulated, and then the successive convex approximation based iterative algorithm was adopted to solve this non-convex optimization problem. Differently, the authors in~\cite{Ma2016Cooperative} proposed a spectrum sharing approach based on a contract-based cooperative technique to obtain an opportunistic spectrum access policy for D2D users while maximizing the cellular network's profit. Specifically, the cooperative spectrum trading process between CUs and D2D devices was modeled based on a pre-defined principal-agent approach. Under this approach, the cellular network acts as a principal and offers a power-payment contract to the D2D pair. After that, the D2D pair, which acts as an agent, tries to choose a contract that can maximize its utility function. Based on this design, the authors can derive optimal contracts that the cellular system can offer and obtain the optimal contract-based spectrum sharing policy.

Although demonstrating good performance for spectrum sharing in D2D communication, the applications of these studies may be limited. This is because the existing solutions are mainly based on static optimization techniques which require all information of the system in advance to formulate the problem and obtain the optimal solution. Unfortunately, it is difficult, if not impossible, to obtain complete prior knowledge of the system due to the dynamics and uncertainty of wireless communications and the mobility of users. To deal with this problem, deep reinforcement learning (DRL) has emerged as a promising tool recently~\cite{Budhiraja2021Deep, Xiang2022Multi, Huang2021Deep, Huang2022dynamic, Liang2024Deep}. By observing the system states and outcomes after performing actions, DRL can efficiently learn the dynamics and uncertainty of the system to converge to the optimal spectrum access policy without requiring prior knowledge about CUs, BSs, and D2D users. For example, the authors in~\cite{Huang2021Deep} and~\cite{Huang2022dynamic} developed DRL-based algorithms that help D2D users dynamically access the shared spectrum to maximize the system's sum throughput without using any prior knowledge about the system. With the proposed DRL-based solutions, D2D transmitters can learn the characteristics of CUs and the BS to decide when they can access the shared spectrum. Simulation results then showed that the proposed DRL-based approaches can achieve a better sum throughput compared to a baseline based on BS cooperation.

\subsection{Motivations and Main Contributions}
Unfortunately, the aforementioned studies and others in the literature may not work well when there are a large number of CUs operating in the shared spectrum. Under such scenarios, the shared spectrum can be always heavily occupied for transmissions of CUs, resulting in low communication efficiency for D2D users and/or frequent collisions for CUs. To address this essential issue, in this paper, we propose to use the ambient backscatter communication (AmBC) technology~\cite{Liu2013Ambient, Huynh2018Ambient, Hoang2020Ambient} to help D2D users transmit data even when the shared spectrum is occupied by CUs. The key fundamental of the AmBC technology is that it can transmit information by just backscattering ambient RF signals without generating any active signals. For that, when the shared spectrum is occupied, D2D users can switch to the ambient backscatter mode to backscatter RF signals generated from the BS to transmit data to their receivers. The backscattered signals from D2D users do not introduce any noticeable interference to CUs since there is no active signal as demonstrated in~\cite{Liu2013Ambient}. It is worth noting that the AmBC technology has been integrated into D2D communication in a few studies~\cite{Lu2018Wireless, Gao2022Cooperative}. However, dynamic spectrum access for D2D communication is not considered in these works. In particular, the authors in~\cite{Lu2018Wireless} proposed a new self-sustainable communication paradigm for D2D communication by using ambient backscatter communications and then theoretically analyze the average throughput, energy outage probability, and coverage probability of the proposed hybrid D2D communication system. The authors in~\cite{Gao2022Cooperative} instead focused on designing passive relays with ambient backscatter communications in wireless-powered D2D systems.

In addition, current DRL-based spectrum access approaches may take a very long time to converge to the optimal policy, especially in complex problems with high-dimensional state spaces. Moreover, DRL-based approaches usually require large deep neural network architectures to efficiently learn highly dynamic and complex environments, resulting in long training time and thus may not be applicable for applications with low latency requirements~\cite{Hoang2023Deep}. Inspired by the quantum superposition and quantum entanglement principles, we develop a quantum reinforcement learning (RL)-based solution to obtain the optimal dynamic spectrum access policy for D2D communications. Instead of using conventional deep neural networks, our proposed solution employs a parametrized quantum circuit to approximate the optimal dynamic spectrum access policy for D2D devices. Through extensive simulations, we demonstrate that the proposed quantum RL approach can achieve a faster convergence rate with much fewer trainable parameters compared to existing DRL methods. It is worth noting that the proposed quantum RL approach can efficiently run on classical computers by using the TensorFlow Quantum and Cirq libraries~\cite{Broughton2020TensorflowQuantum}. As far as we know, this is the first study to take into account quantum RL and the AmBC technology for dynamic spectrum access in D2D communications. Our main contributions can be summarized in the following.
\begin{figure*}[!]
	\centering
	\includegraphics[scale=0.45]{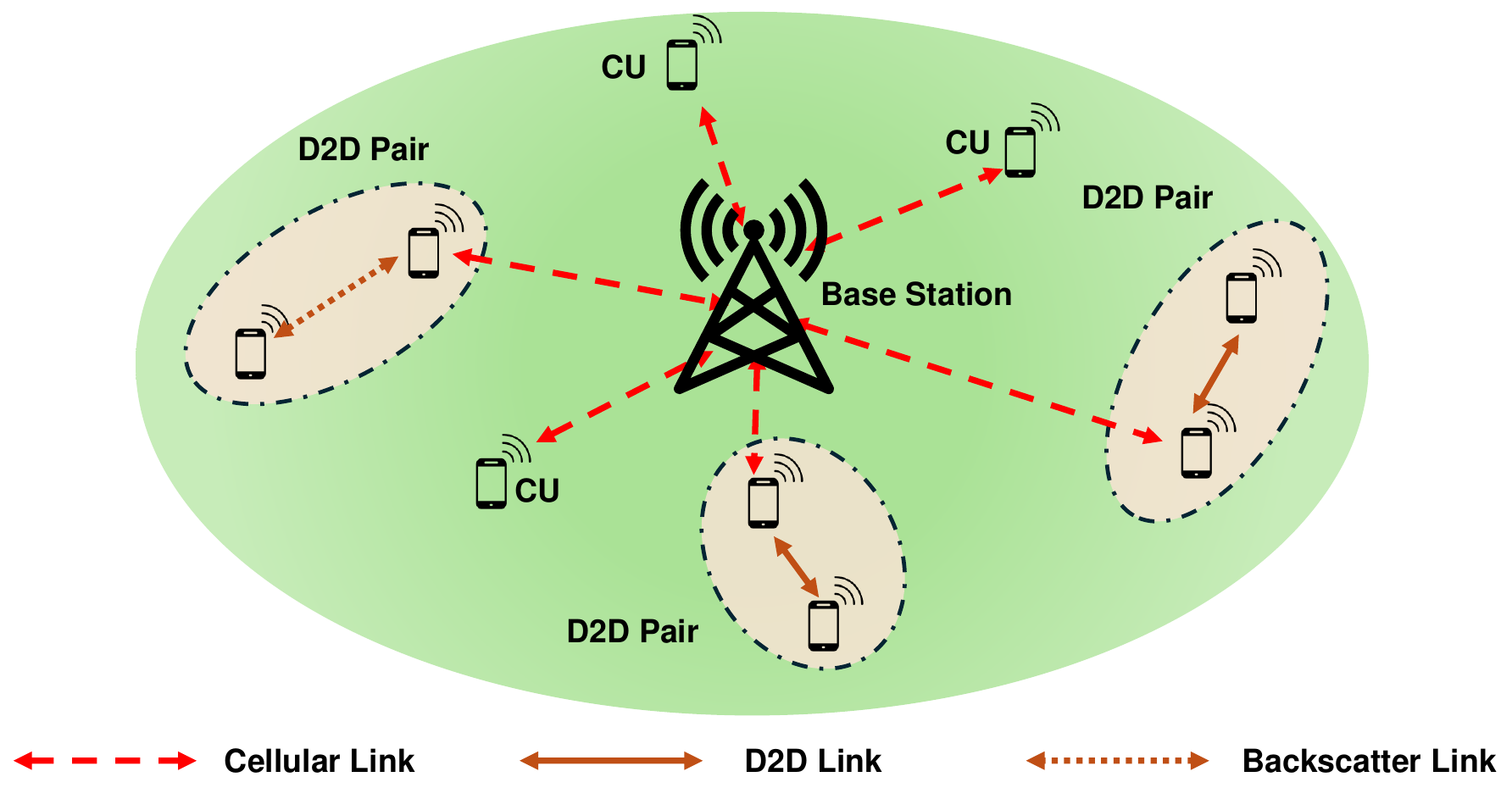}
	\caption{System model of D2D-enabled cellular networks.}
	\label{fig:system model}
\end{figure*}

\begin{itemize}
\item We propose to integrate the AmBC technology into D2D devices for data transmissions by backscattering ambient RF signals, e.g., signals generated by the BS, when the shared spectrum is occupied by CUs. With this new design, D2D devices still can maintain their communications and avoid collisions with CUs even when the shared spectrum is always busy.

\item We use the Markov decision process to model the dynamics and uncertainty of the system caused by user mobility and wireless environments. Then, we develop a DRL-based approach, namely deep Q-learning, to find the optimal spectrum access policy, i.e., stay idle, access the shared spectrum and perform active transmissions, or backscatter the BS's signal for transmissions, to maximize the long-term average throughput of D2D users.

\item To further improve the convergence rate and reduce the training complexity of our proposed solution, we develop a novel quantum RL approach to learn the environment's properties more efficiently and quickly. In particular, a parametrized quantum circuit is used to approximate the optimal policy instead of conventional deep neural networks. With the quantum circuit, the proposed algorithm can leverage the quantum superposition and quantum entanglement principles to better handle the high-dimensional system state space and achieve better learning and system performance than that of the conventional DRL algorithm.

\item Finally, an extensive evaluation is provided to verify the effectiveness of our proposed solution. In particular, with the AmBC technology, D2D users can transmit their data even when the shared spectrum is occupied, resulting in better communication performance. In addition, the proposed quantum RL approach can significantly improve the convergence rate with fewer training parameters and much lower training time compared to existing solutions based on DRL.
\end{itemize}

The rest of the paper is organized as follows. The considered system model, including the D2D and ambient backscatter channel models, is presented in Section~\ref{sec:system_model}. Section~\ref{sec:problem_formulation} introduces our proposed MDP framework to capture the dynamics and uncertainty of the considered system. After that, Section~\ref{sec:quantum} provides the fundamentals of deep RL and our proposed quantum RL algorithms. Section~\ref{sec:performance} then presents the performance analysis in several scenarios. Finally, Section~\ref{sec:conclusion} concludes our work.

\section{System Model}
\label{sec:system_model}

In this paper, we consider a D2D-enabled cellular network which consists of multiple CUs and a BS, as illustrated in Fig.~\ref{fig:system model}. In the considered network, wireless devices can communicate with each other without routing packets through the cellular network by using D2D communications. Without the loss of generality, we assume that D2D nodes and CUs can share the spectrum resources in a time-splitting manner~\cite{Zhao2019mobile, Huang2022dynamic}. Similar to~\cite{Huang2022dynamic}, the time slot allocation among CUs is orthogonal. We then define $p_\mathrm{access}$ as the probability that a CU accesses the shared spectrum in each time slot. As mentioned, D2D nodes can also access this shared spectrum using the same time slots\footnote{Since the main aim of this paper is developing a novel dynamic spectrum access framework for D2D devices to efficiently share the same spectrum with CUs, we consider the case that different D2D pairs are allocated with different resource blocks. In fact, several solutions have been proposed to deal with this issue by using clustering techniques. More details can be found in~\cite{Gharbieh2019Self} and~\cite{Sheng2015On}.}. However, D2D transmitters may generate interference and disrupt the transmissions of CUs if they access the shared spectrum at the same time. As such, D2D nodes must find appropriate time slots to communicate to avoid introducing interference to CUs. In addition, as studied in~\cite{Huang2022dynamic}, a CU suffers from less interference from D2D nodes if it is near the BS. Thus, we define $d_\mathrm{protected}$ as the maximum distance between CUs and the BS that allows CUs to communicate without being affected by interference from D2D nodes. We define $p_\mathrm{protected}$ as the probability that a CU is in the secure area, i.e., its distance to the BS is less than $d_\mathrm{protected}$.

In the future wireless networks, a large number of diverse wireless devices can share the same communication spectrum in a dense area. As such, it is very challenging for D2D devices to opportunistically access the shared spectrum to transmit their data. To further improve the spectrum usage efficiency and improve the system throughput, we propose to use the AmBC technology~\cite{Huynh2018Ambient, Hoang2020Ambient} for D2D nodes. In particular, with the ambient backscatter circuit, D2D nodes can communicate with each other by simply reflecting or absorbing ambient RF signals, i.e., cellular signals in our considered system. To do that, each D2D node (e.g., mobile phones or IoT devices) is equipped with an RF switch that can switch between two different loads to reflect or absorb cellular signals. In the reflecting state, D2D nodes can transmit bits ``1''. In contrast, D2D nodes can transmit bits ``0'' in the absorbing state. In this way, the AmBC technology can help D2D nodes transmit information even when the current time slot is occupied by CUs without introducing any transmission disruptions to CUs. We then denote $\mathcal{C} \triangleq \{c: c\in \{0, 1, 2, 3\}\}$ as the set of the channel states. In particular, $c = 0$ when the channel is idle, $c=1$ when the channel is only used by a CU, $c=2$ when the channel is used by a D2D pair, and $c=3$ when the channel is used by a CU and a D2D pair for their transmissions. Without loss of generality, we assume that the considered system is saturated in which all the nodes always have information to transmit~\cite{Yu2019Deep}. Each node can only transmit information at the beginning of its assigned time slot and must finish its transmission within this time slot. In this paper, we aim to obtain an optimal and dynamic spectrum access policy for D2D nodes to maximize their average throughput. In the following, we present the channel models of D2D communications and ambient backscatter communications. Overall, with the AmBC technology, the D2D transmitter simply absorbs or reflects ambient RF signals for its transmissions instead of generating active signals like D2D communications.

\subsection{D2D Communication Channel Model}
Practically, D2D links can be modeled by using a probabilistic path-loss model, including a non-line-of-sight (NLOS) link and a line-of-sight (LOS) link~\cite{Huang2022dynamic}. As described in~\cite{pathloss}, the LOS path loss can be expressed as follows:
\begin{equation}
L_\mathrm{LOS} = 16.9\log_{10}(d_\mathrm{tr}) + 32.8 +20.0\log_{10}(f),
\end{equation}
where $d_\mathrm{tr}$ is the distance between the D2D transmitter (D2D-Tx) and the D2D receiver (D2D-Rx), and $f$ is the center frequency. Similarly, the NLOS path loss can be expressed as follows:
\begin{equation}
L_\mathrm{NLOS} = 40.0\log_{10}(d_\mathrm{tr}) + 79.0 +30.0\log_{10}(f).
\end{equation}
The average path loss of the D2D link in dB then can be calculated as follows:
\begin{equation}
L = P_\mathrm{LOS} L_\mathrm{LOS} + (1-P_\mathrm{LOS}) L_\mathrm{NLOS},
\end{equation}
where $P_\mathrm{LOS}$ and $(1-P_\mathrm{LOS})$ represent the probability of having an LOS link and an NLOS link, respectively. Based on the distance between the D2D-Tx and the D2D-Rx, $P_{\mathrm{LOS}}$ can be calculated as follows\cite{pathloss}:
\begin{equation}
P_{\mathrm{LOS}}	=	\left\{	\begin{array}{ll}
	1 &	d_\mathrm{tr} \leq 4, \\
	\exp(-(d_\mathrm{tr}-4)/3) &	4 < d_\mathrm{tr} \leq 60, \\
	0 &	d_\mathrm{tr} \geq 60.
\end{array}	\right.
\end{equation}

Let $P_d$ be the transmit power of the D2D-Tx. The average signal power received by the D2D-Rx can be expressed as follows:
\begin{equation}
P_r[\mathrm{mW}] = 10^{(P_d[\mathrm{dBm}]-L[\mathrm{dB}])/10}.
\end{equation}
The D2D connection's achievable transmission rate can be calculated as follows:
\begin{equation}
\label{eq:d2d_rate}
C_\mathrm{d} = W\log_2(1 + P_r / P_n),
\end{equation}
where $W$ is the bandwidth of the channel and $P_n$ is the noise power.

\subsection{Ambient Backscatter Communication Channel Model}
In this paper, we adopt the AmBC technology to allow D2D nodes to communicate with each other even when the spectrum is occupied by CUs. In particular, when the BS is communicating with a CU in the share spectrum, D2D nodes can reflect or absorb the cellular signals to transmit their information without generating active RF signals, and thus do not interrupt the transmissions of the CU. Interested readers can refer to~\cite{Huynh2018Ambient} and~\cite{Hoang2020Ambient} for more information about the design, principles, and circuits of the AmBC technology. In the following, we provide the channel model and formulate the achievable rate of ambient backscatter communications.

Specifically, the RF signals sent from the BS can be expressed as~\cite{Li2018Adaptive}:
\begin{equation}
x(n) = \sqrt{P_t}s(n),
\end{equation}
where $P_t$ is the transmit power of the BS, and $s(n)$ with $\mathbb{E}(|s(n)|^2) = 1$ is the transmitted signal at the $n$-th symbol interval of the BS. Then, the ambient RF signal received at the D2D-Tx can be formulated as follows:
\begin{equation}
y_\mathrm{dt}(n) =  h_\mathrm{st}x(n) = \sqrt{P_t} h_\mathrm{st} s(n),
\end{equation}
where $h_\mathrm{st}$ is the channel coefficient between the BS and the D2D-Tx. Let $\alpha$ denote the reflection coefficient at the D2D-Tx and $b(n)$ denote the D2D-Tx's signal at the $n$-th symbol interval, we then can express the backscatter signal from the D2D-Tx as follows:
\begin{equation}
x_\mathrm{dt}(n) = \sqrt{\alpha} b(n) y_\mathrm{dt} = \sqrt{\alpha P_t} h_\mathrm{st} b(n) s(n).
\end{equation}

The backscattered signal received at the D2D-Rx then can be expressed as follows:
\begin{equation}
y_\mathrm{tr}(n) = h_\mathrm{tr} x_\mathrm{dt}(n) = \sqrt{\alpha P_t} h_\mathrm{st} h_\mathrm{tr} b(n) s(n),
\end{equation}
where $h_\mathrm{tr}$ is the channel coefficient between the D2D-Tx and the D2D-Rx. The BS's signal received at the D2D-Rx can be also expressed as follows:
\begin{equation}
y_\mathrm{dr}(n) = h_\mathrm{dr} x(n) = \sqrt{P_t} h_\mathrm{dr} s(n),
\end{equation}
where $h_\mathrm{dr}$ is the channel coefficient between the BS and the D2D-Rx. Then, the total received signal at the D2D-Rx can be expressed as:
\begin{equation}
\begin{aligned}
	y(n) & = y_\mathrm{tr}(n) + y_\mathrm{dr} + N(n) \\
	&= \sqrt{\alpha P_t} h_\mathrm{st} h_\mathrm{tr} b(n) s(n) + \sqrt{P_t} h_\mathrm{dr} s(n) + N(n),
\end{aligned}
\end{equation}
where $N(n)$ is the Gaussian noise at the D2D-Rx with zero mean and variance $\sigma^2$, i.e., $N(n) \sim \mathcal{CN}(0, \sigma^2)$.

Similar to~\cite{Kang2018Riding, Zhuang2020Optimal}, we assume that the D2D-Rx is equipped with the \emph{successive interference cancellation (SIC)} technique to decode the received signals. In particular, the SIC technique, which is a common physical-layer approach, is usually used to handle two or more signals received at a receiver. With SIC, the D2D-Rx can decode the stronger signals, i.e., the BS signal, first, subtract it from the received signals, and then extract the weaker signal, i.e., the backscattered signal, from the residue~\cite{Kang2018Riding}. In this way, the signal-to-noise (SNR) ratio at the D2D-Rx $\gamma$ can be expressed as:

\begin{equation}
\gamma = \frac{\alpha P_t g_\mathrm{st} g_\mathrm{tr}}{\sigma^2},
\end{equation}
where $g_\mathrm{st} = |h_\mathrm{st}|^2$ and $g_\mathrm{tr} = |h_\mathrm{tr}|^2$ are the channel power gains from the BS to the D2D-Tx and from the D2D-Tx to the D2D-Rx, respectively~\cite{Li2018Adaptive}. Finally, the achievable backscatter rate can be calculated as follows~\cite{Zhuang2020Optimal}:

\begin{equation}
\label{eq:backscatter_rate}
C_\mathrm{b} = W \log_2(1 + \gamma) = W \log_2(1 + \frac{\alpha P_t g_\mathrm{st} g_\mathrm{tr}}{\sigma^2}).
\end{equation}

In this work, we aim to maximize the average throughput of D2D nodes by obtaining an optimal dynamic spectrum access policy. In particular, a D2D node needs to decide when it can access the shared spectrum and which communication technology, i.e., D2D communication or AmBC, it can use to transmit data. Given the fact that CUs can randomly arrive at the BS's coverage area and access the shared spectrum, it is very challenging to obtain the optimal dynamic spectrum access policy. To address this practical problem, in the following, we present a novel quantum RL approach that can dynamically and intelligently learn the system's properties and CUs' behaviors to approximate an optimal spectrum access policy with a superior convergence rate compared to existing solutions.

\section{Problem Formulation}
\label{sec:problem_formulation}
We use the Markov decision process (MDP) to formulate the optimization problem in order to address the dynamic and uncertain nature of the system under consideration. In particular, a tuple $<\mathcal{S}, \mathcal{A}, r>$, where $\mathcal{S}$ denotes the state space, $\mathcal{A}$ represents the action space, and $r$ denotes the immediate reward function, is theoretically used to define an MDP.

\subsection{State Space}
Practically, the channel state can only be accurately observed at the end of a time slot after the agent makes an action at the beginning of the time slot. As such, our system state space includes the channel state at the previous time slot. In addition, to help the agent learn the CU's behaviors and the system's properties more efficiently, we also consider the chosen action and the location of the CU in the previous time slot. As discussed in Section~\ref{sec:system_model}, the achievable rates of D2D communications and ambient backscatter communications depend greatly on the distance between D2D nodes as well as the distance from the D2D-Tx to the BS. For this reason, these two factors are also included in our system state space. Given the above, $\mathcal{S}$ can be formally defined as follows:

\begin{equation}
\mathcal{S} \triangleq \{s: s \in \{a, c, p, d_\mathrm{dt}, d_\mathrm{tr}\}, \forall a \in \mathcal{A}, \forall c \in \mathcal{C}, \forall p\in \{0,1\}\},
\end{equation}
where $a$ denotes the previous action taken by the agent, $c$ denotes the previous channel state, $p$ denotes if the CU in the previous time slot is in the BS's secure area (i.e., $p = 1$) or otherwise (i.e., $p=0$), $d_\mathrm{dt}$ is the distance between the D2D-Tx to the BS, and $d_\mathrm{tr}$ is the distance between the D2D-Tx and the D2D-Rx. In this way, the system state at time slot $t$ can be formally expressed as $s_{t} = (a_{t-1}, c_{t-1}, p_{t-1}, d_\mathrm{dt}, d_\mathrm{tr}) \in \mathcal{S}$.

\subsection{Action Space}
As mentioned in Section~\ref{sec:system_model}, the D2D-Tx can choose to perform D2D communications or ambient backscatter communications to transmit data to the D2D-Rx. In addition, it can choose to stay idle. For that, the action space can be formally defined as follows:
\begin{equation}
\mathcal{A} \triangleq \{a: a \in \{0, 1, 2\}\},
\end{equation}
where $a = 0$ if the D2D-Tx stays idle, $a = 1$ if the D2D-Tx performs D2D communications, and $a = 2$ if the D2D-Tx chooses to leverage the AmBC technology for its transmissions.

\subsection{Immediate Reward}
In this work, our aim is to maximize the average throughput of D2D devices while minimizing the interference introduced to CUs. For that, the immediate reward after the D2D-Tx takes action $a_t$ at state $s_t$ can be defined as follows:
\begin{equation}
\label{eq:reward_function}
r_t(s_t, a_t)	=	\left\{	\begin{array}{ll}
	0, &	\mbox{if } a_t = 0, \\
	C_\mathrm{d}, &	\mbox{if } a_t = 1 \mbox{ and there is no active CU}, \\
	C_\mathrm{d}, &	\mbox{if } a_t = 1 \mbox{ and there is an active CU}\\
	& \mbox{in the secured area}, \\
	P, & \mbox{if } a_t = 1 \mbox{ and there is an active CU}\\
	& \mbox{outside the secured area},\\
	C_\mathrm{b}, &	\mbox{if } a_t = 2.\\
\end{array}	\right.
\end{equation}

Specifically, when the D2D-Tx chooses to stay idle (i.e., $a_t = 0$) the immediate reward will be 0. If the D2D-Tx transmits its information by D2D communications (i.e., $a_t = 1$) and there is no CU actively using the shared spectrum, the immediate reward will be $C_\mathrm{d}$ as defined in~(\ref{eq:d2d_rate}). In addition, if there is an active CU but this CU is in the secured area, then the immediate reward is also $C_\mathrm{d}$. In contrast, if this CU is outside the secured area, the immediate reward will be $P$, which is a very low value to ensure that the agent will avoid actions that introduce interference to CUs. Finally, if the D2D-Tx chooses to use the ambient backscatter to transmit data (i.e., $a_t = 2$), the immediate reward will be $C_\mathrm{b}$ as defined in~(\ref{eq:backscatter_rate}).

\subsection{Long-term Average Throughput Optimization Formulation}
The long-term average throughput maximization problem for D2D devices given the formulated MDP can be expressed as follows:
\begin{eqnarray}
	\label{eq:average_reward}
	\max_\pi	& &	{\mathcal{R}}(\pi)	=	\lim_{T \rightarrow \infty} \frac{1}{T} \sum_{t=1}^{T} {\mathbb{E}} \left( r_t (s_t, \pi(s_t)) \right),
\end{eqnarray}
where $\pi$ is a spectrum access policy which is a mapping from state $s_{t}$ to action $a_{t}$, $r_t (s_t, \pi(s_t))$ represents the immediate reward received after taking an action following policy $\pi$ at state $s_t$, and $\mathcal{R}(\pi)$ is the average long-term throughput under policy $\pi$. It is worth noting that with our formulated MDP, $\mathcal{R}(\pi)$ is independent from the initial state $s_0$ of the system as the underlying Markov chain is irreducible~\cite{Hoang2023Deep}. Thus, the optimal dynamic spectrum access policy exists and can be obtained.

\section{Quantum Reinforcement Learning}
\label{sec:quantum}

\subsection{Preliminaries of Reinforcement Learning and Deep Q-learning}
\label{subsec:deepq}
To efficiently solve the optimization problem in Section~\ref{sec:problem_formulation}, RL algorithms can be adopted. Among RL algorithms, Q-learning and deep Q-learning (DQL) are the most common algorithms adopted in the field of communications and networking~\cite{Hoang2023Deep}. Fundamentally, Q-learning and DQL algorithms are designed based on the Bellman equation to perform simple value iteration updates to approximate the Q-values for all state-action pairs.

More specifically, the Q-learning algorithm stores the Q-values of all possible state-action combinations in a Q-table. Given state $s_t$, the algorithm can take action $a_t$ based on the $\epsilon$-greedy approach. Under this approach, a random action can be taken with probability $\epsilon$, and the action that has the highest Q-value given the current state will be chosen with probability $(1-\epsilon)$. Then, the immediate reward and next state of the system will be observed after taking action $a_t$. With this observation, the Q-learning algorithm can update the Q-value of state-action pair $(s_t, a_t)$ as in~(\ref{eq:qfunction})~\cite{watkins1992q}, where $r_t(s_t, a_t)$ represents the immediate reward after taking action $a_t$ at state $s_t$, $\gamma \in [0,1)$ is the discount factor that determines the future reward's importance, and $\tau_t$ represents the algorithm's learning rate.
\begin{equation}
\label{eq:qfunction}
\begin{aligned}
	\mathcal{Q}_{t+1}(s_t,a_t) = \mathcal{Q}_t(&s_t,a_t) + \tau_t \Big [ r_t(s_t, a_t)\\
	& + \gamma\max_{a_{t+1}} \mathcal{Q}_t(s_{t+1}, a_{t+1})- \mathcal{Q}_t(s_t,a_t)\Big ],
\end{aligned}
\end{equation}
Theoretically, a small value of $\gamma$ indicates that short-term rewards are more important. Differently, the algorithm prefers actions that maximize long-term rewards when $\gamma$ approaches 1. Since we aim to maximize the long-term average throughput of D2D devices, $\gamma$ will be set at high values, e.g., $0.9$. Differently, the learning rate denotes the significance of new observations over existing observations. To enable a stable learning process, the learning rate of Q-learning can be set at small values, e.g., 0.1. Based on~(\ref{eq:qfunction}), the Q-learning algorithm can gradually update the Q-values for all state-action pairs in the system to converge to the optimal spectrum access policy. However, the Q-learning algorithm has several limitations. In particular, it requires the state space to be discrete in order to build the Q-table which may result in information loss. In addition, the Q-learning algorithm cannot work well with high-dimensional state spaces due to the high complexity nature of the Q-table. As such, the performance of Q-learning is limited in practical problems in communications and networking as demonstrated in the literature~\cite{Hoang2023Deep,van2019jam}.

To solve this practical issue of Q-learning, Google DeepMind introduced DQL, a combination of Q-learning and deep neural networks, in 2015~\cite{mnih2015human}. The key idea is using a deep neural network, called deep Q-network, to approximate the Q-values instead of using the Q-table. With the power in handling high-dimensional data, the deep Q-network can help DQL to learn the environment more effectively compared to Q-learning. In addition, the state space now can be continuous, thus avoiding the information loss problem as in Q-learning due to the discretization process. To improve the learning efficiency and stability of the algorithm, two mechanisms, namely \textit{experience replay} and \textit{quasi-static target Q-network}, are adopted. In the \textit{experience replay} mechanism, a memory pool will be used to store previous experiences/observations. The algorithm then can randomly take a number of experiences to train the deep Q-network in each training step. With the \textit{quasi-static target Q-network} mechanism, a separate deep Q-network is employed, called target Q-network. The parameters of this target Q-network are slowly but frequently updated by copying the parameters of the deep Q-network. The target Q-network is then used to approximate target Q-values. As a result, the target Q-values and the estimated Q-values are not correlated, and thus minimizing overestimation and stabilizing the algorithm.

The loss function of the DQL algorithm can be expressed as follows:
\begin{equation}
\label{eq:lossfunction}
\begin{aligned}
	L_i(\theta_i)=&\mathbb{E}_{(s,a,r,s')\sim U(\mathbf{D})}\bigg[ \bigg( r + \\
	&\gamma\max_{a'}\mathcal{Q}(s',a';\theta_i^-) -\mathcal{Q}(s,a;\theta_i)\bigg)^2\bigg],
\end{aligned}
\end{equation}
where $\theta_i$ denotes the parameters of the deep Q-network at time slot $i$ and $\theta_i^-$ represents the parameters of the target Q-network at time slot $i$. To update these parameters, the \textit{Stochastic Gradient Descent} algorithm and its extensions~\cite{Goodfellow2016Deep} can be adopted to minimize the loss in~(\ref{eq:lossfunction}). In this work, Adam optimizer will be used to minimize the loss and update the deep Q-network's parameters.

\subsection{Proposed Quantum Reinforcement Learning}
\begin{figure*}[!]
\centering
\includegraphics[scale=0.42]{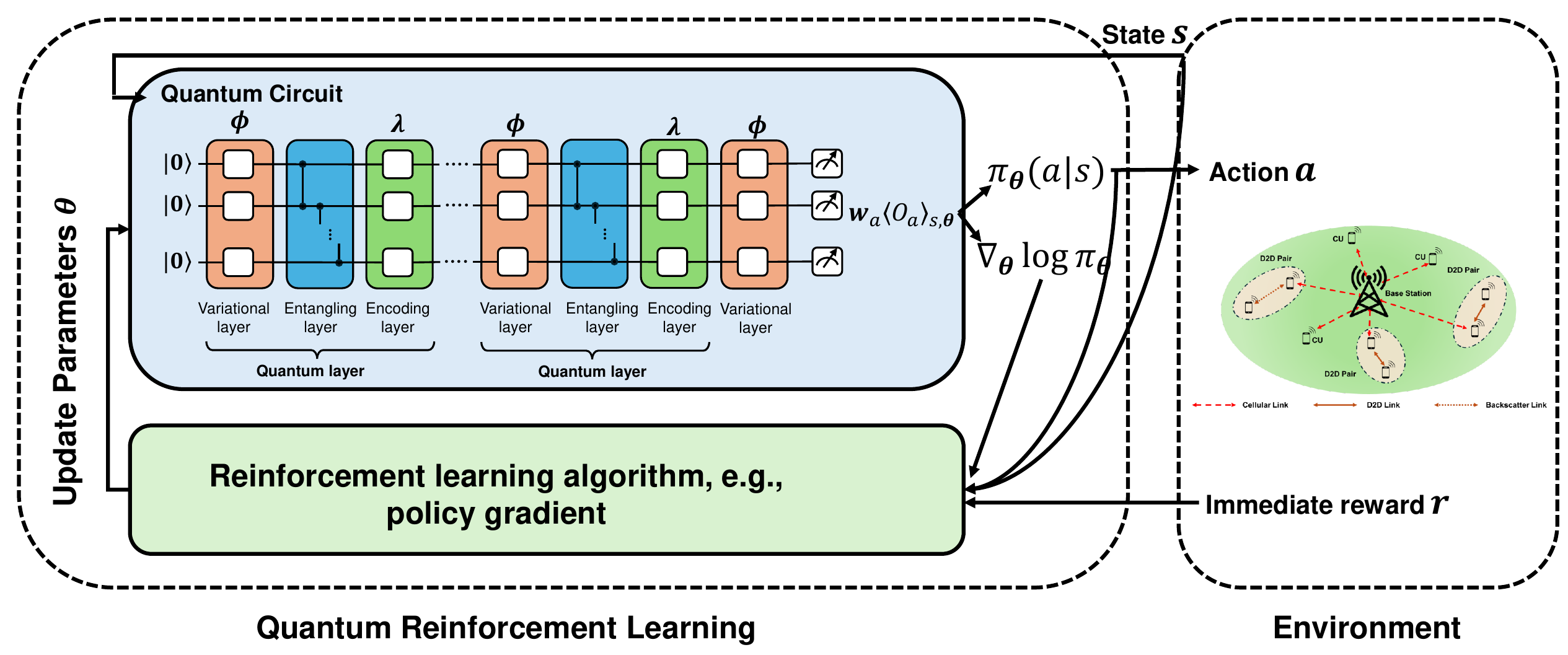}
\caption{Quantum RL architecture~\cite{Jerbi2021Parametrized}.}
\label{fig:quantum_rl}
\end{figure*}

Although DRL algorithms such as DQL presented in Section~\ref{subsec:deepq} have been widely adopted for solving problems in wireless communications with good performance, they still have several limitations that may hinder their applications in future wireless communication systems. In particular, since DRL uses deep neural networks for approximating the policy, the training time can be long, especially in complex problems with high-dimensional state spaces. This may make DRL inapplicable in highly dynamic communication systems where the system conditions can be changed quickly and in real-time scenarios. To overcome this issue, in this paper, we develop a novel quantum RL algorithm to efficiently approximate the optimal spectrum access policy with a much faster convergence rate compared to DRL. The key idea is to use a quantum circuit for approximating the optimal policy instead of using deep neural networks, as illustrated in Fig.~\ref{fig:quantum_rl}. In this way, the \emph{quantum entanglement} and the \emph{quantum superposition} properties of quantum computing can be utilized to speed up the learning process of RL, resulting in a high convergence rate~\cite{Narottama2023Quantum, Zaman2023Quantum}. Next, we present our proposed quantum RL approach in detail.

\subsubsection{Parametrized Quantum Circuit}

\begin{figure}[h]
	\centering
	\includegraphics[scale=0.45]{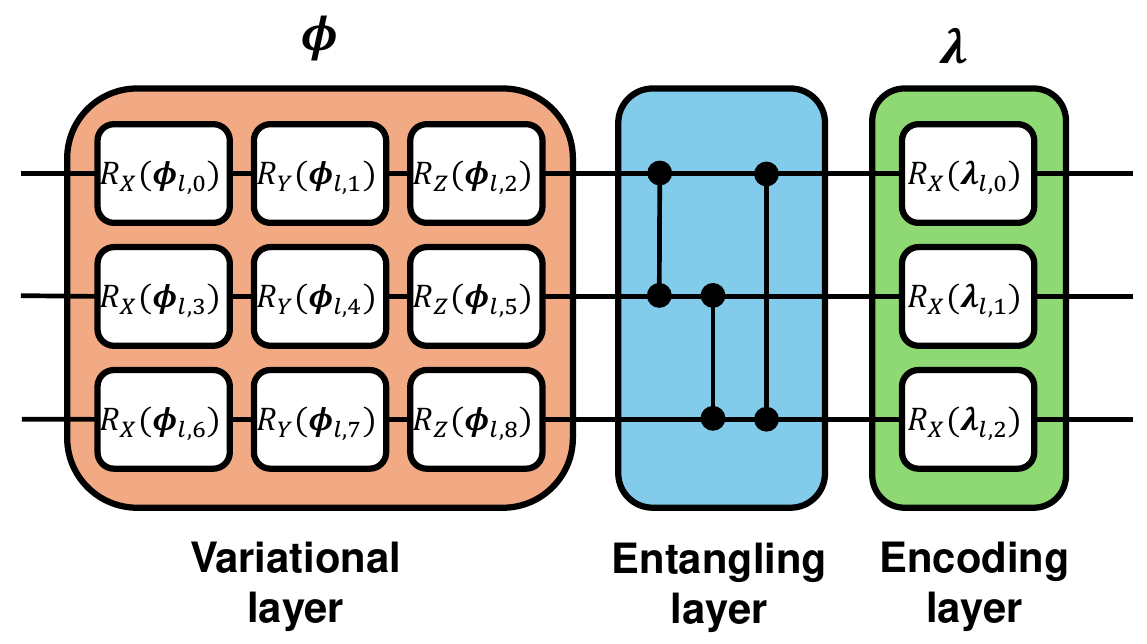}
	\caption{Architecture of a quantum layer.}
	\label{fig:pqc}
\end{figure}

In our proposed quantum RL approach, we employ a parametrized quantum circuit (PQC) to approximate the optimal policy. In particular, the proposed PQC takes the system state $s$ as its input and outputs a vector of expectation values. Then, this output vector can be processed to obtain the policy $\pi(a|s)$, i.e., policy-based method, or the approximated Q-values $\mathcal{Q}(s,a)$, i.e., value-based method. In this work, we use policy-based RL to train the proposed PQC. As studied in\cite{Jerbi2021Parametrized} and~\cite{Skolik2022Quantum}, the PQC design has a significant impact on the learning performance. Among several factors, the data encoding strategy plays an essential role when designing a PQC. Based on theoretical analysis in~\cite{Salinas2020Data} and~\cite{Schuld2021Effect}, the data re-uploading technique outperforms others by constructing a highly-expressive learning model. In particular, instead of encoding data only once by the variational layer, the data re-uploading method aims to encode data in several encoding layers interlayed with variational layers. For that reason, we use the data re-uploading technique to build our PQC in this paper. As illustrated in Fig.~\ref{fig:quantum_rl}, the proposed PQC first uses a variational layer to handle the input state vector by performing single-qubit rotations with trainable angles $\bm{\phi} \in [0, 2\pi]^{|\bm{\phi}|}$. The outputs of this variational layer are processed by an entangling layer which consists of several 2-qubit controlled Z (CZ) gates. After that, the outputs of the entangling layer are fed into an encoding layer with trainable input-scaling parameters $\bm{\lambda} \in \mathbb{R}^{|\bm{\lambda}|}$. Next, the outputs of this encoding layer are processed by another variational layer. A PQC can have multiple entangling layers, encoding layers, and variational layers. We define a \emph{quantum layer} to consist of a variational layer, an entangling layer, and an encoding layer. Fig.~\ref{fig:pqc} illustrates the detailed architecture of a \emph{quantum layer}. It is worth noting that the last layer in the proposed PQC is the variational layer. This is because the variational layer can help to increase the circuit's expressibility by representing more quantum states than the encoding layer~\cite{Salinas2020Data}. In addition, the encoding layer can learn only a sine function of its input, resulting in a less efficient training process~\cite{Coelho2024VQC}. The estimated policy $\pi(a|s)$ then can be obtained by observing the expectation values of the last variational layer. The number of expectation values is the number of actions in the proposed MDP, i.e., $|\mathcal{A}|$. Similar to~\cite{Salinas2020Data} and~\cite{Broughton2020TensorflowQuantum}, we also employ trainable weight $w_a \in \bm{w}$ to augmented the expectation value of action $a$. By optimizing the trainable weights $\bm{\phi}$, $\bm{\lambda}$, and $\bm{w}$, we can learn the environment and approximate the dynamic spectrum access policy for D2D devices. In the following, we present the details of how the proposed PQC can process the system state (i.e., classical data) to quantum states.

As we feed the system state to our PQC, the number of qubits $n$ is the number of features in the state space $\mathcal{S}$ (i.e., $n=5$ in our paper). With $n$ qubits, the PQC can be represented by a $2^n$-dimensional complex Hilbert space $\mathcal{H} = (\mathbb{C}^2)^{\otimes n}$~\cite{Jerbi2021Parametrized}. The quantum state of the PQC can be defined by a vector $|\psi\rangle \in \mathcal{H}$ with unit norm $\langle\psi|\psi\rangle = 1$. The quantum superposition can then be defined as follows:
\begin{equation}
|\psi\rangle = \alpha|0\rangle + \beta|1\rangle,
\end{equation}
where $\alpha$ and $\beta$ are the complex coefficients with $|\alpha|^2 + |\beta|^2 = 1$. As illustrated in Fig.~\ref{fig:pqc}, in the PQC, there are several quantum gates which are defined as unitary operations performing on $\mathcal{H}$. In the variational layers, we consider the single-qubit Pauli gates $X$, $Y$, and $Z$ whose matrices are defined as follows:

\begin{equation}
X = \begin{pmatrix}
	0 & 1\\
	1 & 0
\end{pmatrix},
Y = \begin{pmatrix}
	0 & -i\\
	i & 0
\end{pmatrix},
Z = \begin{pmatrix}
	1 & 0\\
	0 & -1
\end{pmatrix}.
\end{equation}
In the variational layers, classical data is encoded to quantum states by using single-qubit rotations corresponding to the single-qubit gates as follows:
\begin{equation}
\begin{aligned}
	R_X(\phi_{l,i}) = \exp\big(-i\frac{\phi_{l,i}}{2}X\big),\\
	R_Y(\phi_{l,i}) = \exp\big(-i\frac{\phi_{l,i}}{2}Y\big), \\
	R_Z(\phi_{l,i}) = \exp\big(-i\frac{\phi_{l,i}}{2}Z\big),
\end{aligned}
\end{equation}
where $\phi_{l,i} \in \bm{\phi}, \forall l, i,$ is the rotation angle of the $l$-th variational layer. Given the above, state $\bm{s}$ (i.e., classical data) can be encoded into a quantum state as follows~\cite{Coelho2024VQC}:
\begin{equation}
|\bm{s}\rangle = \bigotimes_{k=0}^{n-1}R(\phi_{l,i} \times s_k) |0_k\rangle,
\end{equation}
where $n$ is the number of dimensions of the state space, i.e., number of qubits, $R \in \{R_X, R_Y, R_Z\}$, $\phi_{l,i}$ is the rotation angle corresponding to the single-qubit rotation $R$, and $|0_k\rangle$ is the $k$-th qubit initialized in state $|0\rangle$~\cite{Coelho2024VQC}. The tensor product $\bigotimes$ is used to combine individual qubits into an entangled quantum state.

In the encoding layers, we only consider the X gate together with trainable scale parameters $\bm{\lambda}$. For the entangling layers, we use the 2-qubit controlled Z gates which can be defined as:
\begin{equation}
CZ = \begin{pmatrix}
	1 & 0 & 0 & 0\\
	0 & 1 & 0 & 0\\
	0 & 0 & 1 & 0\\
	0 & 0 & 0 & -1
\end{pmatrix}.
\end{equation}
As mentioned, the output of the last variational layer, defined as observable $\mathcal{O}$, can be used to obtain the policy $\pi(a|s)$. However, as studied in~\cite{Jerbi2021Parametrized}, the policy obtained from the PQC may not have a direct adjustable greediness which can affect the exploration and exploitation of the RL process. To address this problem, the softmax activation function can be used to generalize the expected values $\langle P_a \rangle_{s, \bm{\theta}}$ of projection $P_a$, with $\bm{\theta} = (\bm{\phi}, \bm{\lambda, \bm{w}})$.

\subsubsection{Quantum Reinforcement Learning with Policy Gradient Approach}
Given a state $s$ as the input of the proposed PQC, the quantum state corresponding to the unitary $U(s, \bm{\phi}, \bm{\lambda})$ can be expressed as $|\psi_{s, \bm{\phi}, \bm{\lambda}}\rangle = U(s, \bm{\phi}, \bm{\lambda})|0^{\otimes n}$, where $n$ is the number of qubits which is the number of features in the system state space defined in Section~\ref{sec:problem_formulation}, i.e., $n=5$ in our considered dynamic spectrum access problem. Given this quantum state, the policy associated with parameters $\bm{\theta}$ can be defined as follows~\cite{Jerbi2021Parametrized}:
\begin{equation}
\pi_{\bm{\theta}}(a|s) = \frac{e^{\xi \langle \mathcal{O}_a \rangle_{s, \bm{\theta}}}}{\sum_{a'}e^{\xi \langle \mathcal{O}_{a'} \rangle_{s, \bm{\theta}}}},
\end{equation}
where $\xi \in \mathbb{R}$ is an inverse-temperature parameter~\cite{Jerbi2021Parametrized} and $\bm{\theta} = (\bm{\phi}, \bm{\lambda, \bm{w}})$ is the set of trainable angles, trainable input-scaling parameters, and trainable weights defined above. $\langle \mathcal{O}_a \rangle_{s, \bm{\theta}}$ is the expectation value observed from the output of the proposed PQC and can be expressed as follows:
\begin{equation}
\langle \mathcal{O}_a \rangle_{s, \bm{\theta}} = \Big \langle \psi_{s, \bm{\phi}, \bm{\lambda}} \big \vert\sum_i w_{a,i} H_{a,i} \big \vert \psi_{s, \bm{\phi}, \bm{\lambda}} \Big \rangle,
\end{equation}
where $H_{a,i}$ is the weighted Hermitian operator corresponding to action $a$ at iteration $i$. Similar to~\cite{Jerbi2021Parametrized}, we simplify $H_{a,i}$ to the tensor product of Pauli matrices on computational basis states, i.e., $|00\rangle, |01\rangle,|10\rangle$, and $|11\rangle$. By updating the parameters $\bm{\theta}$ of the proposed PQC, we can approximate the optimal policy $\pi^*(a|s)$. As mentioned, in this work, we consider a simple RL approach based on the policy-based mechanism to optimize $\bm{\theta}$ as shown in Algorithm~\ref{quantum_rl}.

\begin{algorithm}
\caption{Quantum Reinforcement Learning based D2D Spectrum Access}
\label{quantum_rl}
\begin{algorithmic}[1]
	\State Initialize an experience batch $\mathcal{B}$ to the size of $B$.
	\State Initialize the proposed PQC with parameters $\bm{\theta}= (\bm{\phi}, \bm{\lambda}, \bm{w})$.
	\For{\textit{t=1 to T}}
	\State \multiline{Given state $s_t$, performing action $a_t$ based on the current policy and observe immediate reward $r_t$ and next state $s_{t+1}$}
	\State \multiline{Store transition $(s_t, a_t, r_t, s_{t+1})$ in the experience batch $\mathcal{B}$}
	\If{$modulo(t, B) == 0$}
	\State \multiline{Perform a gradient descent step on the loss function in~(\ref{eq:loss_function_quantum}) with respect to the parameters $\bm{\theta} = (\bm{\phi}, \bm{\lambda}, \bm{w})$ of the PQC}
	\EndIf
	\EndFor
\end{algorithmic}
\end{algorithm}

In particular, the algorithm aims to interact with the environment, i.e., the AmBC-assisted D2D system in our work, to update the PQC's parameters $\bm{\theta}$ through the gradient descent on the value function $V_{\pi_{\bm{\theta}}}(s_0) = \mathbb{E}_{\pi_{\bm{\theta}}}\Big[\sum_{t=0}^{T}\gamma^t r_t\Big]$. The loss function then can be derived based on the policy gradient theorem as follows\cite{Broughton2020TensorflowQuantum}:
\begin{equation}
\label{eq:loss_function_quantum}
\mathcal{L}(\bm{\theta}) = -\frac{1}{B} \sum_{s, a, r, s' \in \mathcal{B}} \Big[\sum_{t=1}^{T}\log(\pi_{\bm{\theta}}(a_t|s_t))\sum_{t'=2}^{T-t}\gamma^{t'}r_{t+t'}\Big],
\end{equation}
where $\mathcal{B}$ is a batch of experiences collected after interacting with the environment and $B=|\mathcal{B}|$ is the size of the experience batch. This loss function can be minimized by performing gradient descent on (\ref{eq:loss_function_quantum}), and thus updating the PQC's parameters $\bm{\theta}$. In this work, we do not perform the training at every iteration. Instead, the algorithm only performs training after every $|\mathcal{B}|$ iterations. This not only can reduce the computational complexity of the algorithm but also can stabilize the learning process.

\subsection{Parameters and Complexity Analysis}
As mentioned, quantum RL can not only learn the environment more efficiently by leveraging the superposition of quantum computing but also significantly reduce the number of training parameters compared to the traditional deep neural network architecture. In this section, we analyze the complexity of the proposed quantum RL algorithm and the DQL algorithm in terms of the number of trainable parameters.

\subsubsection{Deep Q-learning}
The DQL algorithm usually uses the standard multilayer perceptron architecture, which consists of multiple fully connected layers, for its deep neural networks. For that, we assume that the DQL algorithm employs a deep neural network that consists of input layer $L_0$, $N$ fully connected hidden layers $(L_1, \ldots, L_N)$, and output layer $L_{N+1}$. It is well known that the trainable parameters of the deep neural network are weights and bias. As such, the number of trainable parameters for connecting layer $L_i$ and $L_{i+1}$ can be calculated as $|L_i||L_{i+1}| + |L_{i+1}| = (|L_i|+1)|L_{i=1}|$, where $|L_i|$ is the number of neurons in layer $L_i$. Based on this, the total trainable parameters of the deep neural network can be formulated as follows:
\begin{equation}
\begin{aligned}
	P_\mathrm{DNN} &= (|L_0| + 1)|L_1| + \ldots + (|L_N| + 1)|L_\mathrm{N+1}|\\
	&= \sum_{i=0}^{N} (|L_i| + 1)|L_\mathrm{i+1}|.
\end{aligned}
\end{equation}

It is well known that a typical deep neural network requires several hidden layers with 32 to 128 neurons in each layer for good training performance, resulting in a few thousand trainable parameters. For complicated problems, the number of parameters can be much larger as more layers and neurons are needed to efficiently learn the complex environments.

\subsubsection{Quantum Reinforcement Learning}
Similar to the above, we will formulate the computational complexity of the proposed PQC in terms of trainable parameters in the following. In particular, in the variational layer, we use three gates, i.e., $R_X$, $R_Y$, and $R_Z$, for qubit rotations of each qubit. As such, with $n$ qubits, the size of the variational layer is $3n$. Differently, in the encoding layer, we only use gate $R_X$ for each qubit. For that, the size of the encoding layer is $n$. In summary, the number of trainable parameters in \emph{quantum layers} can be calculated as follows:
\begin{equation}
P_\mathrm{quantum \: layers} = N(3n+n) = 4Nn,
\end{equation}
where $N$ is the number of \emph{quantum layers}. It is worth noting that there is a variational layer at the end of the PQC. The size of this final variational layer is also $3n$. Recall that for the expectation value of each action, we use a trainable weight to better distinguish the potential actions and their impacts on the learning process. As such, the number of these trainable weights will be the number of actions in the action space, i.e., $|\bm{w}| = |\mathcal{A}|$.

Given the above, the total number of trainable parameters of the proposed PQC can be formulated as follows:
\begin{equation}
P_\mathrm{quantum} = P_\mathrm{quantum \: layers} + 3n + |\mathcal{A}| = (4N+3)n + |\mathcal{A}|.
\end{equation}
It is clear that the complexity of the proposed PQC is relatively small compared to that of the DQL algorithm. For example, in our considered problem, we have 5 dimensions in the state space, and thus $n=5$. With three quantum layers, the total trainable weights of the proposed PQC is only $(4\times3 + 3) \times 5 + 3 = 78$. Meanwhile, the deep neural network in the DQL algorithm may need thousands of parameters for good performance. For example, a deep neural network consists of 2 hidden layers with 32 neurons each has 1,347 trainable parameters. In the next section, we will show that with a significantly smaller number of parameters, quantum RL can achieve much better performance compared to the DQL algorithm. With low complexity, our proposed quantum RL can be deployed and efficiently run on D2D devices, e.g., mobile phones and vehicles. With resource-constrained devices such as wireless sensors and IoT devices, the proposed approach may not be applicable. Instead, it can be deployed at the cluster head or gateway, and they can distribute the dynamic spectrum access policy to other devices in their clusters or networks.

\section{Performance Evaluation}
\label{sec:performance}
This section provides an extensive performance evaluation of the quantum RL approach compared with other baselines in various scenarios. In particular, we first present the parameter setting for the considered system and also the proposed quantum RL approach. Then, we analyze the simulation results in terms of convergence rate, running time, and average throughput in different scenarios.

\subsection{Parameter Setting}
Unless otherwise stated, the probability that a CU accesses the shared spectrum in each time slot $p_\mathrm{access}$ is set at 0.8. As mentioned in Section~\ref{sec:system_model}, in our considered system, there is a secure area near the BS where the interference from D2D devices does not affect the signal reception of CUs if they access the shared spectrum at the same time. For that, we set the probability that a CU is in the secure area $p_\mathrm{protected}$ at 0.2. In this paper, we consider the case that D2D devices are randomly placed in the considered area. Without loss of generality, the distance between a D2D-Tx and the BS $d_\mathrm{st}$ is randomly generated from 100 meters to 1,000 meters and the distance between a D2D-Tx and its receiver $d_\mathrm{tr}$ is randomly generated from 10 meters to 100 meters. The transmit power of the BS $P_t$ is set at 40 dBm. The transmit power of D2D devices $P_d$ is set at 23 dBm and the noise power $P_n$ is set at -114 dBm~\cite{Huang2022dynamic}. The bandwidth and the center frequency are set at 20 MHz and 2 GHz, respectively. $P$ is set at -100 to prevent actions that cause interference to CUs. For the ambient backscatter circuit, we set the reflection efficiency $\alpha$ at 0.6. Similar to~\cite{Lyu2018The}, we set $g_\mathrm{st} = \frac{A_e}{4 \pi d_\mathrm{st}^2}$ and $g_\mathrm{tr} = \frac{A_e}{4 \pi d_\mathrm{tr}^2}$, where $A_e$ is the effective area of the antennas which is 0.0086 $m^2$.

For the DQL algorithm, we use standard parameters used widely in the literature~\cite{Hoang2023Deep}. In particular, we employ 2 hidden layers with the size of $32$ neurons each. The activation function is \emph{tanh}. The learning rate is set at $0.01$ and the optimizer is Adam. For the $\epsilon$-greedy method, $\epsilon$ is set to 1 at the beginning of training and is gradually reduced to $0.01$ with a decay factor of $0.9999$. The memory pool can store up to $10,000$ experiences and the target update frequency $C$ is set at $5,000$. The batch size $B$ is set at 32. For our proposed quantum RL approach, we use 3 quantum layers. We employ three Adam optimizers to optimize $\bm{\phi}, \bm{\lambda}, \bm{w}$ with the learning rates of 0.05, 0.01, and 0.1, respectively. The TensorFlow Quantum library~\cite{Broughton2020TensorflowQuantum} is used to build the proposed quantum circuit.

In this paper, we compare our proposed quantum RL approaches with three baselines in terms of the average throughput which is calculated by the number of bits the D2D-Tx can transmit in one second. The three baselines used in this paper include the following.

\begin{itemize}
\item \emph{Random}: In this method, the D2D-Tx will randomly choose an action in $\mathcal{A}$ for its transmission. This is used to show the system performance of a non-learning approach.
\item \emph{Greedy}: Under this method, the D2D-Tx always chooses to access the shared spectrum and actively transmit its data. This method is used to show the advantages of the AmBC technology, especially when the shared spectrum is mostly occupied by CUs.
\item \emph{DQL}: In this baseline, we use the proposed deep Q-learning algorithm to solve the formulated MDP in Section~\ref{sec:problem_formulation}. By comparing with this method, we can better highlight the advantages of our proposed quantum RL approach compared to the state-of-the-art solutions in the literature.
\end{itemize}

\subsection{Simulation Results}
\subsubsection{Convergence and Complexity Analysis}

\begin{figure}[!]
\centering
\includegraphics[scale=0.45]{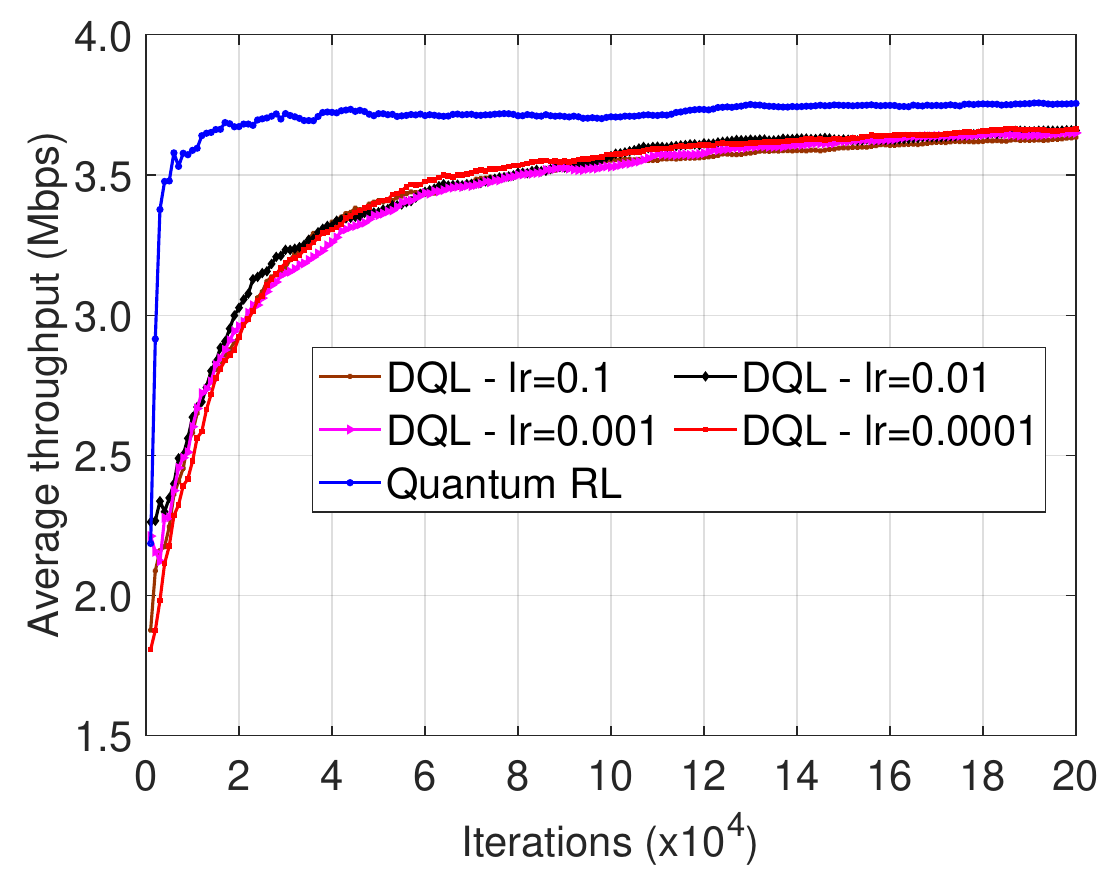}
\caption{Convergence rates of quantum RL and DQL w. layer size of 16.}
\label{fig:convergence_16}
\end{figure}

In Fig.~\ref{fig:convergence_16}, we compare the convergence rate of our proposed quantum RL algorithm with the DQL algorithm under different learning rates. In this scenario, the deep Q-network has two fully connected hidden layers with a size of 16. As can be observed in the figure, the proposed quantum RL algorithm has a much faster convergence rate compared to the DQL algorithm. Specifically, the proposed quantum RL approach can converge at an average throughput of around 3.73 Mbps after 50,000 training steps while the DQL algorithm cannot converge to that performance after 200,000 iterations. This demonstrates the effectiveness of using the proposed PQC to learn the considered environment, thanks to the quantum superposition principle.

\begin{figure}[!]
\centering
\includegraphics[scale=0.45]{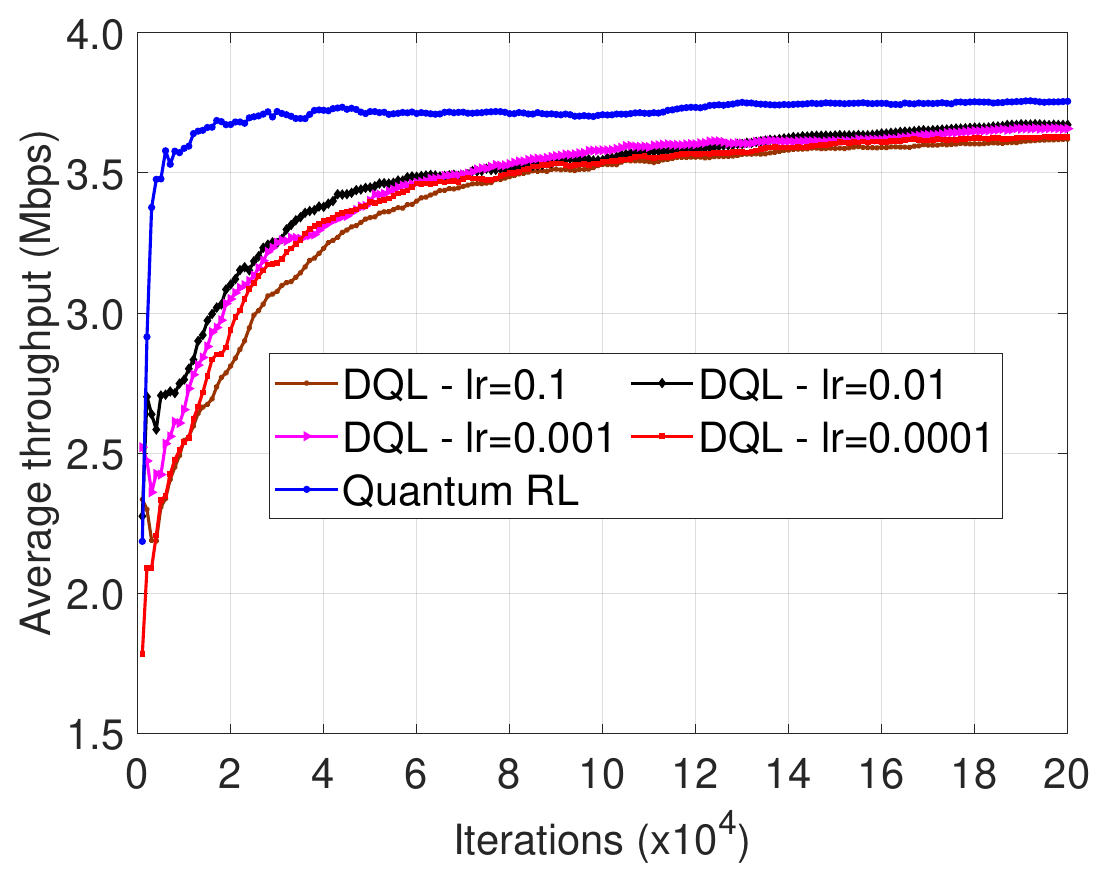}
\caption{Convergence rates of quantum RL and DQL w. layer size of 32.}
\label{fig:convergence_32}
\end{figure}

Next, we increase the size of the hidden layers in the deep Q-network to 32 and compare the convergence rates of deep Q-learning and our proposed quantum RL method, as shown in Fig.~\ref{fig:convergence_32}. As can be observed, with more neurons in each hidden layer, the DQL algorithm can learn faster. However, it still cannot converge to a better policy after 200,000 iterations, while our proposed quantum RL algorithm can quickly learn the environment and obtain a much better average throughput after 20,000 iterations only. It is worth noting that the DQL algorithm achieves the best performance with the learning rate of 0.01. Therefore, in the following simulations, we set the learning rate of the DQL algorithm to 0.01.

\begin{table}[!]
\caption{Complexity Analysis}
\label{tab:complexity}
\centering
\begin{tabular}{|>{\centering\arraybackslash}p{1.8cm}|>{\centering\arraybackslash}p{1.5cm}|>{\centering\arraybackslash}p{1.8cm}|>{\centering\arraybackslash}p{1.5cm}|}
	\hline
	\textbf{Method} & \textbf{No. of parameters} & \textbf{Running time per iteration (seconds)} & \textbf{Average throughput (Mbps)} \\ \hline
	\hline
	DQL-16  & 419  & 0.05281 & 3.40\\
	\hline
	DQL-32  & 1,347  & 0.06254 & 3.46 \\
	\hline
	Quantum RL-1  &  38 & 0.00436 & 3.64 \\
	\hline
	Quantum RL-3  & 78  & 0.00499 & 3.76\\
	\hline
	Quantum RL-5  & 118 & 0.00589  & 3.72\\
	\hline
\end{tabular}
\end{table}

In Table~\ref{tab:complexity}, we compare the number of trainable parameters, the running time of each training step, and the achieved performance after 50,000 training steps of our proposed solution and the DQL algorithm under different settings. As can be observed, the number of training parameters of the proposed quantum RL is significantly smaller than that of the DQL algorithm, i.e., 38, 78, and 118 parameters with 1 quantum layer, 3 quantum layers, and 5 quantum layers, respectively, compared to 419 and 1,347 parameters with the hidden layer size of 16 and 32, respectively. We then run the proposed quantum RL and DQL algorithms in a standard laptop with a Core i5-1235U CPU and 16 GB of RAM. As can be observed from Table~\ref{tab:complexity}, the proposed quantum RL algorithm also runs about ten times faster than the DQL algorithm, thanks to the use of the PQC. Clearly, with 3 quantum layers, the proposed quantum RL algorithm can achieve the best performance. As such, in the following evaluation, we use 3 quantum layers in the proposed PQC. Both the quantum RL and DQL algorithms will be evaluated after 50,000 training steps.

\subsubsection{Performance Evaluation}

\begin{figure}[!]
\centering
\includegraphics[scale=0.45]{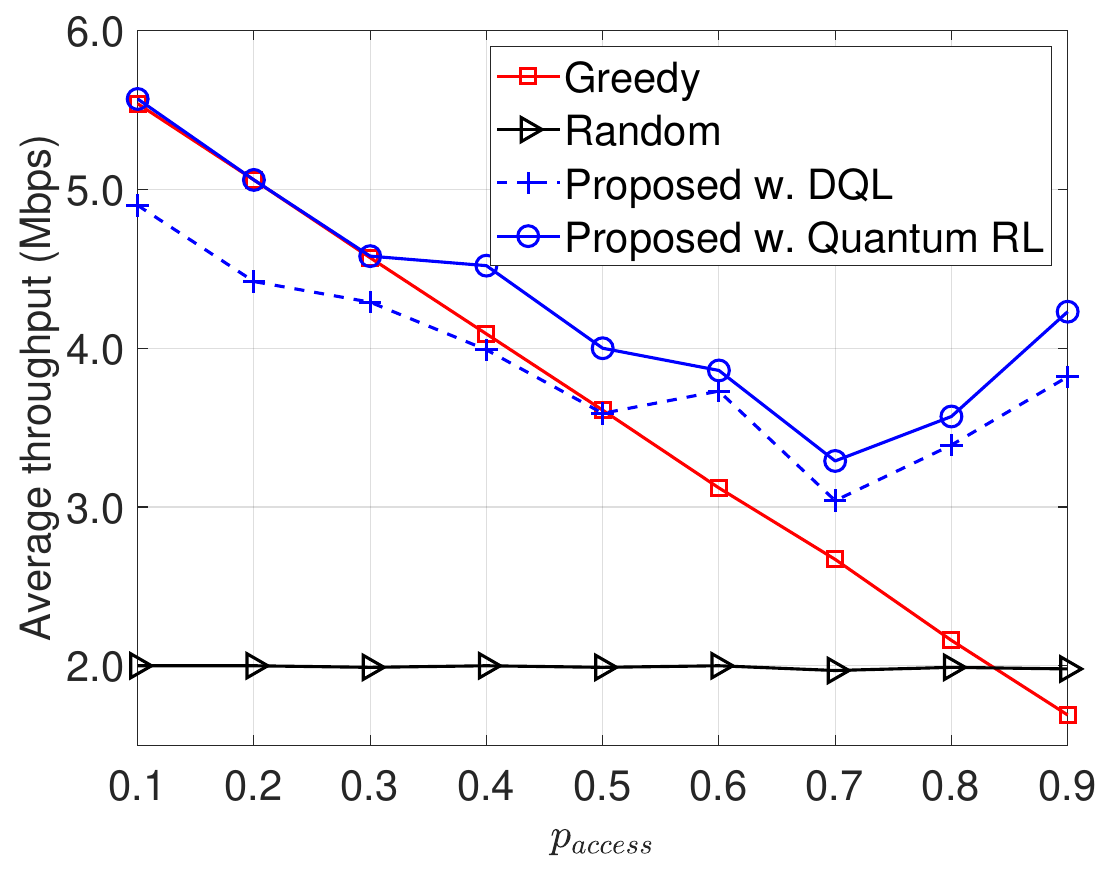}
\caption{Average throughput vs. CU access probability $p_\mathrm{access}$.}
\label{fig:vary_p_access}
\end{figure}

In Fig.~\ref{fig:vary_p_access}, we vary the probability that a CU accesses the shared spectrum in each time slot and compare the average throughput of all the methods. As can be seen, the proposed quantum RL algorithm achieves the best performance compared to other baselines. This is because the proposed PQC can help the algorithm to learn the considered environment more efficiently, resulting in better performance. It is worth noting that when $p_\mathrm{access}$ is high, the performance gap becomes bigger. This stems from the fact that when the shared spectrum is frequently occupied by CUs, D2D transmitters cannot actively transmit their data to D2D receivers. In contrast, with the AmBC technology, our proposed solution still allows D2D transmitters to backscatter information to their receivers, resulting in better throughput. More importantly, when $p_\mathrm{access}$ is higher than $0.7$, the average throughput obtained by the quantum RL and DQL methods increases. This is because, when CUs are likely to access the shared spectrum to communicate with the BS, D2D devices have more chances to backscatter RF signals generated from the BS. This demonstrates the effectiveness of the AmBC technology. Since the DQL algorithm has a slow convergence rate, its performance fluctuates and is inferior to the proposed quantum RL approach.

\begin{figure}[!]
	\centering
	\includegraphics[scale=0.45]{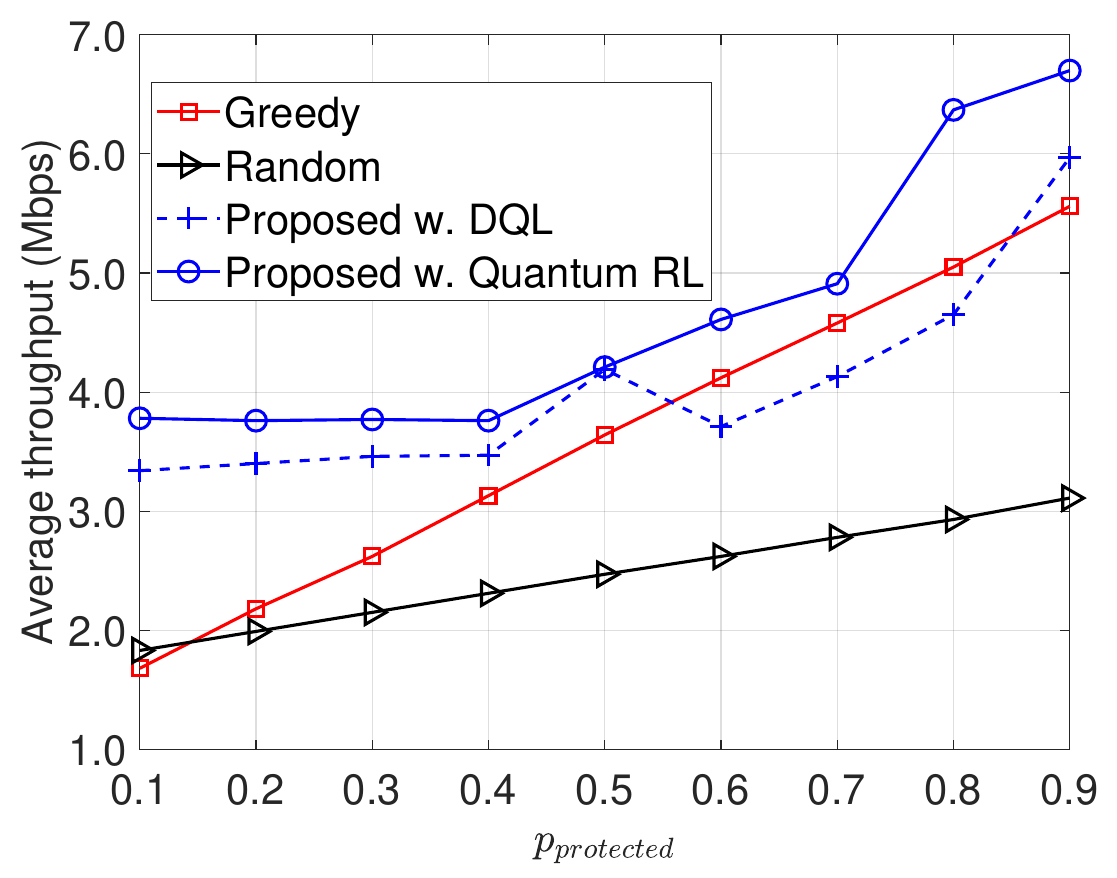}
	\caption{Average throughput vs. probability that CU is in secure area $p_\mathrm{protected}$.}
	\label{fig:vary_p_secured}
\end{figure}

Finally, we vary the probability that a CU is in the secure area and compare the average throughput obtained by all methods, as illustrated in Fig.~\ref{fig:vary_p_secured}. It can be observed that when $p_\mathrm{protected}$ increases, the average throughput of all methods increases. The reason is that when CUs are likely to be in the secure area, they are less vulnerable to interference caused by D2D devices when they use the shared spectrum, resulting in a better throughput. However, in all cases, the proposed quantum RL approach achieves the best average throughout, thanks to the use of the quantum proposition principle.

\section{Conclusion}
\label{sec:conclusion}
In this paper, we have proposed a dynamic spectrum access approach for D2D communications by leveraging the AmBC technology and quantum RL. In particular, by using the AmBC technology, D2D devices can transmit their information even when CUs are accessing the shared spectrum by simply backscattering the RF signals sent from the base station. This approach is particularly effective when the shared spectrum is usually busy which will be a common situation in future dense heterogeneous wireless networks. In addition, it is challenging to obtain the optimal spectrum access policy for D2D devices given the dynamics and uncertainty of the system due to the nature of wireless communication as well as the mobility and behaviors of mobile users. To address this problem, we have developed a quantum RL that can efficiently and quickly learn the environment to approximate the optimal policy by leveraging the quantum superposition principle. Extensive simulations have demonstrated that our proposed solution not only can improve the average throughput of D2D communications but also can quickly learn the environment with significantly fewer training parameters compared to the state-of-the-art methods.

\appendices


\begin{thebibliography}{1}
\bibliographystyle{IEEEtran}

\bibitem{Asadi2014Survey}
A.~Asadi, Q.~Wang, and V.~Mancuso, ``A survey on device-to-device communication in cellular networks,'' \emph{IEEE Communications Surveys \& Tutorials}, vol. 16, no. 4, pp. 1801-1819, Fourth Quarter 2014.

\bibitem{Jiang2016Energy}
Y.~Jiang, Q.~Liu, F.~Zheng, X.~Gao, and X.~You, ``Energy-efficient joint resource allocation and power control for D2D communications,'' \emph{IEEE Transactions on Vehicular Technology}, vol. 65, no. 8, pp. 6119-6127, Aug. 2016.

\bibitem{Ansari20185G}
R.~I.~Ansari, C.~Chrysostomou, S.~A.~Hassan, M.~Guizani, S.~Mumtaz, J.~Rodriguez, and J.~J.~P.~C.~Rodrigues, ``5G D2D networks: Techniques, challenges, and future prospects,'' \emph{IEEE Systems Journal}, vol. 12, no. 4, pp. 3970-3984, Dec. 2018.

\bibitem{Lin2014Spectrum}
X.~Lin, J.~G.~Andrews, and A.~Ghosh, ``Spectrum sharing for device-to-device communication in cellular networks,'' \emph{IEEE Transactions on Wireless Communications}, vol. 13, no. 12, pp.6727-6740, Dec. 2014.

\bibitem{Chu2019Opportunistic}
Z.~Chu, W.~Yu, P.~Xiao, F.~Zhou, N.~A.-Dhahir, A.~U.~Quddus, and R.~Tafazolli, ``Opportunistic spectrum sharing for D2D-based URLLC,'' \emph{IEEE Transactions on Vehicular Technology}, vol. 68, no. 9, pp. 8995-9006, Sept. 2019.

\bibitem{Ma2016Cooperative}
C.~Ma, Y.~Li, H.~Yu, X.~Gan, X.~Wang, Y.~Ren, and J.~J.~Xu, ``Cooperative spectrum sharing in D2D-enabled cellular networks,'' \emph{IEEE Transactions on Communications}, vol. 64, no. 10, pp. 4394-4408, Oct. 2016.

\bibitem{Huang2021Deep}
J.~Huang, Y.~Yang, G.~He, Y.~Xiao, and J.~Liu, ``Deep reinforcement learning-based dynamic spectrum access for D2D communication underlay cellular networks,'' \emph{IEEE Communications Letters}, vol. 25, no. 8, pp. 2614-2618, Aug. 2021.

\bibitem{Huang2022dynamic}
J.~Huang, Y.~Yang, Z.~Gao, D.~He, and D.~W.~K.~Ng, ``Dynamic spectrum access for D2D-enabled Internet of Things: A deep reinforcement learning approach,'' \emph{IEEE Internet of Things Journal}, vol. 9, no. 18, pp. 17793-17807, Sept. 2022.

\bibitem{Liang2024Deep}
Y.-J.~Liang, Y.-C.~Tseng, and C.-W.~Hsieh, ``A deep reinforcement learning-based D2D spectrum allocation underlaying a cellular network,'' \emph{Wireless Networks}, pp. 1-7, May 2024.

\bibitem{Budhiraja2021Deep}
I.~Budhiraja, N.~Kumar, and S.~Tyagi, ``Deep-reinforcement-learning-based proportional fair scheduling control scheme for underlay D2D communication,'' \emph{IEEE Internet of Things Journal}, vol. 8, no. 5, pp. 3143-3156, Mar. 2021.

\bibitem{Xiang2022Multi}
H.~Xiang, Y.~Yang, G.~He, J.~Huang, and D.~He, ``Multi-agent deep reinforcement learning-based power control and resource allocation for D2D communications,'' \emph{IEEE Wireless Communications Letters}, vol. 11, no. 8, pp. 1659-1663, Aug. 2022.

\bibitem{Gao2022Cooperative}
X.~Gao, D.~Niyato, K.~Yang, and J.~An, ``Cooperative scheme for backscatter-aided passive relay communications in wireless-powered D2D networks,'' \emph{IEEE Internet of Things Journal}, vol. 9, no. 1, pp. 152-164, Jan. 2022.

\bibitem{Liu2013Ambient}
V.~Liu, A.~Parks, V.~Talla, S.~Gollakota, D.~Wetherall, and J.~R.~Smith, ``Ambient backscatter: Wireless communication out of thin air,'' in \emph{Proc. ACM SIGCOMM}, Hong Kong, Aug. 2013, pp. 39–50.

\bibitem{Huynh2018Ambient}
N.~V.~Huynh, D.~T.~Hoang, X.~Lu, D.~Niyato, P.~Wang, and D.~I.~Kim, ``Ambient backscatter communications: A contemporary survey,'' \emph{IEEE Communications Surveys \& Tutorials }, vol. 20, no. 4, pp. 2889-2922, Fourth Quarter 2018.

\bibitem{Hoang2020Ambient}
D.~T.~Hoang, D.~Niyato, D.~I.~Kim, N.~V.~Huynh, and S.~Gong, \emph{Ambient Backscatter Communication Networks}, 1st ed. Cambridge,
U.K.: Cambridge Univ. Press, 2020.


\bibitem{Lu2018Wireless}
X.~Lu, H.~Jiang, D.~Niyato, D.~I.~Kim, and Z.~Han, ``Wireless-powered device-to-device communications with ambient backscattering: Performance modeling and analysis,'' \emph{IEEE Transactions on Wireless Communications}, vol. 17, no. 3, pp. 1528-1544, Mar. 2018.



\bibitem{Gharbieh2019Self}
M.~Gharbieh, A.~Bader, H.~ElSawy, H.-C.~Yang, M.-S.~Alouini, and A.~Adinoyi, ``Self-organized scheduling request for uplink 5G networks: A D2D clustering approach,'' \emph{IEEE Transactions on Communications}, vol. 67, no. 2, pp. 1197-1209, Feb. 2019.

\bibitem{Sheng2015On}
M.~Sheng, H.~Sun, X.~Wang, Y.~Zhang, T.~Q.S.~Quek, J.~Liu, and J.~Li, ``On-demand scheduling: Achieving QoS differentiation for D2D communications,'' \emph{IEEE Communications Magazine}, vol. 53, no. 7, p. 162-170, Jul. 2015.



\bibitem{Zhao2019mobile}
G.~Zhao, S.~Chen, L.~Qi, L.~Zhao, and L.~Hanzo, ``Mobile-traffic-aware offloading for energy-and spectral-efficient large-scale D2D-enabled cellular networks,'' \emph{IEEE Transactions on Wireless Communications}, vol. 18, no. 6, pp. 3251-3264, Jun. 2019.

\bibitem{Yu2019Deep}
Y.~Yu, T.~Wang, and S.~C.~Liew, ``Deep-reinforcement learning multiple access for heterogeneous wireless networks,'' \emph{IEEE Journal on Selected Areas in Communications}, vol. 37, no. 6, pp. 1277-1290, Jun. 2019.

\bibitem{pathloss}
``Study on LTE device to device proximity services; radio aspects, release 12,'' 3GPP, Sophia Antipolis, France, Rep. TR 36.843, Mar. 2014.

\bibitem{Li2018Adaptive}
D.~Li and Y.-C.~Liang, ``Adaptive ambient backscatter communication systems with MRC,'' \emph{IEEE Transactions on Vehicular Technology}, vol. 67, no. 12, pp. 12352-12357, Dec. 2018.

\bibitem{Kang2018Riding}
X.~Kang, Y.-C.~Liang, and J.~Yang, ``Riding on the primary: A new spectrum sharing paradigm for wireless-powered IoT devices,'' \emph{IEEE Transactions on Wireless Communications}, vol. 17, no. 9, pp. 6335-6347, Sept. 2018.

\bibitem{Zhuang2020Optimal}
Y.~Zhuang, X.~Li, H.~Ji, H.~Zhang, and V.~C.~M.~Leung, ``Optimal resource allocation for RF-powered underlay cognitive radio networks with ambient backscatter communication,'' \emph{IEEE Transactions on Vehicular Technology}, vol. 69, no. 12, pp. 15216-15228, Dec. 2020.

\bibitem{Hoang2023Deep}
D.~T.~Hoang, N.~V.~Huynh, D.~N.~Nguyen, E.~Hossain, and D.~Niyato, \emph{Deep Reinforcement Learning for Wireless Communications and Networking: Theory, Applications and Implementation}. John Wiley \& Sons, 2023.

\bibitem{watkins1992q}
C.~J.~C.~H.~Watkins and P.~Dayan, ``Q-learning,'' \emph{ Machine Learning}, vol. 8, no. 3–4, pp. 279–292, 1992.

\bibitem{van2019jam}
N.~V.~Huynh, D.~N.~Nguyen, D.~T.~Hoang, and E.~Dutkiewicz, ``Jam me if you can: Defeating jammer with deep dueling neural network architecture and ambient backscattering augmented communications'', \emph{IEEE Journal on Selected Areas in Communications}, vol. 37 , no. 11, pp. 2603-2620, Nov. 2019.

\bibitem{mnih2015human}
V.~Mnih, K.~Kavukcuoglu, D.~Silver, A.~A.~Rusu, J.~Veness, M.~G.~Bellemare, A.~Graves, et al., ``Human-level control through deep reinforcement learning,'' \emph{Nature}, vol. 518, no. 7540, pp. 529-533, Feb. 2015.

\bibitem{Goodfellow2016Deep}
I.~Goodfellow, Y.~Bengio, and A.~Courville, \emph{Deep learning}. MIT press, 2016.	

\bibitem{Narottama2023Quantum}
B.~Narottama, Z.~Mohamed, and S.~Aissa, ``Quantum machine learning for next-G wireless communications: Fundamentals and the path ahead,'' \emph{IEEE Open Journal of the Communications Society}, vol. 4, pp. 2204-2224, Aug. 2023.

\bibitem{Zaman2023Quantum}
F.~Zaman, A.~Farooq, M.~A.~Ullah, H.~Jung, H.~Shin, and M.~Z.~Win, ``Quantum machine intelligence for 6G URLLC,'' \emph{IEEE Wireless Communications}, vol. 30, no. 2, pp. 22-30, Apr. 2023.

\bibitem{Salinas2020Data}
A.~P.-Salinas, A.~C.-Lierta, E.~G.-Fuster, and J.~I.~Latorre, ``Data re-uploading for a universal quantum classifier,'' \emph{Quantum}, vol. 4, pp. 226, Jan. 2020.

\bibitem{Schuld2021Effect}
M.~Schuld, R.~Sweke, and J.~J.~Meyer, ``Effect of data encoding on the expressive power of variational quantum-machine-learning models,'' \emph{Physical Review A}, vol. 103, no. 3, pp. 032430, Mar. 2021.

\bibitem{Jerbi2021Parametrized}
S.~Jerbi, C.~Gyurik, S.~Marshall, H.~Briegel, and V.~Dunjko, ``Parametrized quantum policies for reinforcement learning,'' \emph{Advances in Neural Information Processing Systems (NeurIPS)}, vol. 34, pp. 28362-28375, 2021.

\bibitem{Skolik2022Quantum}
A.~Skolik, S.~Jerbi, and V.~Dunjko, ``Quantum agents in the gym: a variational quantum algorithm for deep q-learning,'' \emph{Quantum 6}, vol. 6, pp. 720, May 2022.

\bibitem{Broughton2020TensorflowQuantum}
M.~Broughton, G.~Verdon, T.~McCourt, A.~J.~Martinez, J.~H.~Yoo, S.~V.~Isakov, P.~Massey \emph{et al.}, ``Tensorflow quantum: A software framework for quantum machine learning,'' arXiv preprint arXiv:2003.02989, 2020.

\bibitem{Lyu2018The}
B.~Lyu, C.~You, Z.~Yang, and G.~Gui, ``The optimal control policy for RF-powered backscatter communication networks,'' \emph{IEEE Transactions on Vehicular Technology}, vol. 67, no. 3, pp. 2804-2808, Mar. 2018.

\bibitem{Coelho2024VQC}
R.~Coelho, A.~Sequeira, and L.~P.~Santos, ``VQC-based reinforcement learning with data re-uploading: performance and trainability,'' \emph{Quantum Machine Intelligence}, vol. 6, no. 2, pp. 53, 2024.

\end{thebibliography}
\end{document}